\documentstyle[12pt]{article}
\begin{document}
\title{Gravitation and Electromagnetism II}
\author{B.G. Sidharth\\
B.M. Birla Science Centre, Adarsh Nagar, Hyderabad - 500 063 (India)}
\date{}
\maketitle
\begin{abstract}
We show how, a non commutative underpinning for spacetime throws up mass,
spin and charge and indicate how it is possible to include QCD effects
also.
\end{abstract}
\begin{flushleft}
Suggested PACS:  04.50.+h  04.20.Gz
\end{flushleft}
\section{Introduction}
In a previous communication\cite{r1} it was suggested that if we consider a
non commutative geometrical underpinnilng to the usual spacetime given by
equation
$$[x,y] = 0(l^2)$$
$$[x, p_x] = \imath \hbar [1+(a/\hbar)^2 p^2_x];$$
$$[t, p_t] = \imath \hbar [1-(a/ \hbar c)^2 p^2_t];$$
\begin{equation}
[x, p_y] = [y, p_x] = \imath \hbar (a/ \hbar)^2 p_xp_y ;\label{e1}
\end{equation}
$$[x, p_t] = c^2[p_{x,} t] = \imath \hbar (a/ \hbar)^2 p_xp_t ;\mbox{etc}.$$
Then it is possible to obtain a unified formulation for electromagnetism and
gravitation. Infact it was shown that using (\ref{e1}), we can deduce that
\begin{equation}
A^\mu = \hbar \Gamma^{\mu v}_v\label{e2}
\end{equation}
where $A^\mu$ is the electromagnetic four potential and $\Gamma^{\mu v}_{\lambda}$ represents
the Christoffel symbols.\\
Equation (\ref{e2}) provides a link between electromagnetism and gravitation, and
infact is mathematically identical to Weyl's original unified formalism\cite{r2},
except that in our case it arises due to spacetime properties given by (\ref{e1}),
and is not adhoc as in the former case.\\
It may be mentioned that the relations (\ref{e1}) arise when there is a minimum
non zero spacetime interval $l$ and $\tau = l/c$, as was demonstrated several
decades ago by Snyder\cite{r3}. Such considerations, once rejected are now
back in the reckoning, for example in the Quantum Superstring theory where also
we have extended particles and the above non commutative geometry\cite{r4}.\\
Let us now pursue the above line of thought a little further.
\section{Mass, Spin and Charge}
Our starting point is the formula for the metric in flat spacetime
\begin{equation}
ds^2 = g_{\mu \nu} dx^\mu dx^\nu\label{e3}
\end{equation}
If we use in (\ref{e3}) the relation (\ref{e1}), after rewriting the product
of the coordinate differentials as a sum of symmetric and anti symmetric terms,
we will get
\begin{equation}
g_{\mu \nu} = \eta_{\mu \nu} + kh_{\mu \nu}\label{e4}
\end{equation}
The first term in (\ref{e4}) denotes the usual flat spacetime metric while
the second term which represents a departure from flat spacetime denotes the
effect of non commutativity, $k$ being a suitable constant.\\
It is interesting to note that such a symmetric-anti symmetric combination for
the coordinate differentials is similar to that of torsional theories\cite{r5},
with however the important difference that in torsional theories the contribution
to the spacetime interval from the anti symmetric term vanishes, where as here,
in view of (\ref{e1}) there is a non zero contribution. This again is because
due to the non commutative geometry, the spacetime points of conventional
theory are ruled out. It must also be noted that if $0(l^2)$ and $0 (\tau^2)$
are neglected in (\ref{e1}) then we are back again with conventional spacetime.\\
We next observe that starting from (\ref{e4}) and remembering that the second
term is small as $l$ and $\tau$ are small, we can with standard arguments
(Cf.\cite{r6}) deduce the linearized general relativistic equation.
\begin{equation}
-\eta^{\mu \nu} \partial_\lambda \partial^\lambda h + \eta^{\mu \nu} \partial_\lambda
\partial_\sigma h^{\lambda \sigma} = - kT^{\mu \nu}\label{e5}
\end{equation}
where the term on the right side of (\ref{e5}) arises because of the fact that
$$
\frac{\partial}{\partial x^\lambda} \frac{\partial}{\partial x^\mu} -
\frac{\partial}{\partial x^\mu} \frac{\partial}{\partial x^\lambda} \quad \mbox{goes}
\quad \mbox{over}\quad \mbox{to} \frac{\partial}{\partial x^\lambda}
\Gamma^\nu_{\mu \nu} - \frac{\partial}
{\partial x^\mu} \Gamma^\nu_{\lambda \nu}
$$
which now does not vanish due to the non commutativity.\\
It can further be shown that using the usual gauge (Cf.\cite{r6}), in view of
(\ref{e2}) we now have
\begin{equation}
\partial_{\mu} h^{\mu \nu} = A^{\nu}\label{e6}
\end{equation}
Also introducing the notation
$$\phi^{\mu \nu} = h^{\mu \nu} - \frac{1}{2} \eta^{\mu \nu} h$$
Equation (\ref{e5}) can be rewritten as
\begin{equation}
\Box \phi^{\mu \nu} = - kT^{\mu \nu}\label{e7}
\end{equation}
It is interesting to note that operating on both sides of (\ref{e7}) with $\partial_\mu$,
we get in view of (\ref{e6}) Maxwell's equations. Equation (\ref{e7}) describes
the spin 2 graviton while the Maxwell equations lead to the spin 1 photon.\\
We now observe that given equation (\ref{e5}) or (\ref{e7}), as is well known
in the linearized theory (Cf. also\cite{r7}) we have, taking for simplicity
$k = 1$,
\begin{equation}
h_{\mu v} = \int
\frac{4T_{\mu v}(t-|\vec x - \vec x'|,\vec x')}{|\vec x - \vec x'|}
d^3x'\label{e8}
\end{equation}
It must be mentioned that equation (\ref{e8}) is valid outside the Schwarzchild
radius, which in any case is the region of our interest. Further velocities
comprable to the velocity of  light are allowed and the stresses $T^{jk}$ and
momentum densities $T^{0j}$ can be comparable to the energy momentum density
$T^{00}$. It can be easily deduced that when $\frac{|\vec x'|}{r} < < 1$,
where $r \equiv | \vec x |$, and in a frame with origin at the centre of mass
and at rest with respect to the particle, and in units in which the Gravitational
constant is unity,
\begin{equation}
m = \int T^{00} d^3 x\label{e9}
\end{equation}
\begin{equation}
S_k = \int \epsilon_{klm} x^{l}T^{m0}d^3 x\label{e10}
\end{equation}
where $m$ is the mass (or approximate mass because of the linear
approximation), and $S_k$ is the angular momentum (Cf.ref.\cite{r7} for details).
We next observe that,
\begin{equation}
T^{\mu v} = \rho u^\mu u^v\label{e11}
\end{equation}
As can be seen from (\ref{e1}), $l$ and $\tau$ are the minimum spacetime
intervals which we take to be at the Compton scale, the Planck scale being a
special case for the Planck mass $\sim 10^{-5}gms$. The reason for this is that
while the Compton scale is the "size" of elementary particles, as is well known,
even in classical physics, this is the limiting case within which we encounter
unphysical negative energies, exactly as in Quantum Theory\cite{r8}. We also use the input
that at the Compton scale, the velocity $u^\imath = c$ the velocity of light
while $u^0 = 1$. Substitution of (\ref{e11}) in (\ref{e10}) now gives, on
using the mean value theorem
\begin{equation}
S_k = c < x^l > \int \rho d^3 x\label{e12}
\end{equation}
As $l \sim \hbar/2mc$ the Compton wavelength, we get from (\ref{e12}) on using
(\ref{e9}),
$$S_k = \hbar/2$$
as required for a Fermion.\\
It is known that the gravitational potential is easily obtained from (\ref{e8})
and (\ref{e9}) and is given by
\begin{equation}
\Phi = - \frac{1}{2}(g^{00} - \eta^{00}) = - \frac{m}{r} + 0
(\frac{1}{r^3})\label{e13}
\end{equation}
(Cf.\cite{r6,r7}).\\
Interestingly, what we have shown is that the particle has mass, spin and charge,
and can be described by the Kerr-Newman metric including the otherwise purely
Quantum Mechanical anomalous gyromagnetic ratio, $g = 2$. For the electron, such a
description throws up a singularity, which however is shielded by the above
unphysical negative energy region within the Compton scale (Cf.\cite{r8}). This
again is symptomatic of the breakdown of the concept of spacetime points, leading
to non commutative geometry.\\
It is also possible to obtain the electromagnetic field from the above by an
alternative method which also enables us to describe QCD effects. Infact using (\ref{e2})
and (\ref{e8}) we get
\begin{equation}
A_0 = 2 \int \eta^{\mu v} \frac{\partial}{\partial t}
[\frac{T_{\mu v}(t-|\vec x - \vec x'|, \vec x')}{|\vec x - \vec x'|}] d^3 x'\label{e14}
\end{equation}
For $|\vec x - \vec x'| = r >> l,$ (\ref{e14}) can be shown to lead to
\begin{equation}
A_0 \sim \frac{\hbar c^3}{r} \int \rho \omega d^3 x'
\sim (mc^3) \frac{mc^2}{r}\label{e15}
\end{equation}
All this has been worked out in detail (Cf.refs.\cite{r8,r9}).\\
Infact if we revert to the usual units, we recover from (\ref{e15}) the well known
electrostatic potential with the well known ratio
$$\frac{e^2}{Gm^2} \sim 10^{40}$$
It is interesting that the same line of argument also leads to the QCD type
of interaction. To see this simply if we expand equation (\ref{e8}) in terms
of a Taylor series about $t$, and this time consider the case where $| \vec x |
\sim |\vec x'|$, we get
\begin{eqnarray}
h_{\mu v} = 4 \int \frac{T_{\mu v}(t,\vec x')}{|\vec x - \vec x'|}d^3 x'+
(\mbox{terms   independent  of}\vec x) + 2 \nonumber \\
\int \frac{d^2}{dt^2} T_{\mu v} (t,\vec x'). |\vec x - \vec x'| d^3 x' +
0(|\vec x - \vec x'|^2)\label{e16}
\end{eqnarray}
Further manipulation (Cf.\cite{r8} for details) yields
\begin{equation}
h_{\mu v} \approx - \frac{\beta M}{r} + 8\beta M
(\frac{Mc^2}{\hbar})^2.r\label{e17}
\end{equation}
$\beta$ being a constant and $M$ being the mass of the particle. (\ref{e17})
already resembles the QCD quark potential\cite{r10} with both the Coulombic
and confining parts. Taking for $M$ the mass of a typical $C$ quark $\sim 1.8Gev$,
the ratio of the coefficients of the $r$ term and the $\frac{1}{r}$ term as
obtained from (\ref{e17}) is $\sim \frac{1}{\hbar^2}(Gev)^2$ as in the case of
QCD\cite{r10}.\\
These considerations can be refined further (Cf.ref.\cite{r11,r12,r13}) and
infact we are able to get not only a characterisation of a QCD potential but
also an explanation for the peculiar characteristics of quarks such as their
one third fractional charge and their handedness.

\end{document}